\documentclass[fp,twocolumn]{jpsj3}
\usepackage{graphicx}
\usepackage{bm}
\usepackage{mathrsfs}

\title{Disentangling Orbital Magnetic Susceptibility with Wannier Functions}
\author{Toshikaze Kariyado$^1$, Hiroyasu Matsuura$^2$, and Masao Ogata$^{2,3}$}
\inst{$^1$ International Center for Materials Nanoarchitectonics, National Institute for Materials Science, Tsukuba, Japan\\ $^2$ Department of Physics, Graduate School of Science, University of Tokyo, Tokyo, Japan\\ $^3$ Trans-scale Quantum Science Institute, University of Tokyo, Tokyo, Japan}

\abst{
Orbital magnetic susceptibility involves rich physics such as interband effects despite of its conceptual simplicity. In order to appreciate the rich physics related to the orbital magnetic susceptibility, it is essential to derive a formula to decompose the susceptibility into the contributions from each band. Here, we propose a scheme to perform this decomposition using the modified Wannier functions. The derived formula nicely decomposes the susceptibility into intraband and interband contributions, and from the other aspect, into itinerant and local contributions. The validity of the formula is tested in a couple of simple models. Interestingly, it is revealed that the quality of the decomposition depends on the degree of localization of the used Wannier functions. The formula here complements another formula using Bloch functions, or the formula derived in the semiclassical theory, which deepens our understanding of the orbital magnetic susceptibility and may serve as a foundation of a better computational method. The relationship to the Berry curvature in the present scheme is also clarified.
}

\begin{document}

\maketitle

\section{Introduction}
The orbital magnetic susceptibility is one of the fundamental responses from electrons as charged particles. In solids, it is sensitive to band structures \cite{Peierls:1933tv} and affected by the structure of Bloch wave functions in momentum space \cite{PhysRevB.91.214405,PhysRevB.94.134423}. Within the semiclassical interpretation, the band structure is reflected in normal velocity, while the wave function structure is reflected in anomalous velocity as Berry curvature. Therefore, experimental measurements of the susceptibility is useful in extracting information of electronic band structure. For instance, it can be used to detect singular band structures in solids such as Dirac electrons \cite{Wehrli:1968uz,doi:10.1143/JPSJ.28.570,PhysRevB.103.115117,PhysRevB.104.035113,tada2021quantum}. In theory, despite its conceptual simplicity, i.e., the magnetic susceptibility is essentially just the lowest-order magnetic moment in external magnetic field, the formula for the orbital magnetic susceptibility is very involved as shown in the early studies \cite{PhysRev.89.633,PhysRevLett.2.150,HEBBORN1960105,PhysRev.126.1636,ROTH1962433,HEBBORN1964741,PhysRev.136.A803,doi:10.1143/JPSJ.20.520}. It was only when Fukuyama combined the Luttinger-Kohn representation \cite{PhysRev.97.869} and the Green's function technique, that we had a gauge invariant compact formula for the orbital magnetic susceptibility \cite{10.1143/PTP.45.704}.

The Fukuyama formula involves Green's function, velocity operator, and trace over all the states. When we apply this formula to a certain model and calculate the orbital magnetic susceptibility numerically, truncation of the trace at some finite number of states is unavoidable. Naively, this truncation by collecting finite number of energy bands sounds reasonable. However, it is not as straightforward as one might think, because it is generically not possible to diagonalize the Green's function and the velocity operator simultaneously. In order to have a good formula, good in the sense that it is useful in numerics and easy to comprehend, we have to handle the offdiagonal matrix elements with great care. Indeed, if we start from the Fukuyama formula in a band basis, we have to make full use of sum rules to isolate the celebrated Landau-Peierls susceptibility from the correction terms \cite{doi:10.7566/JPSJ.84.124708}, where the correction terms are mostly from the local contributions such as the atomic diamagnetism or the van Vleck paramagnetism. Physically, the importance of the offdiagonal matrix elements is from the fact that the orbital magnetic susceptibility is obtained by the second order perturbation that involves virtual hoppings to other bands. There is a big difference between completely neglecting the other bands from the beginning and taking account of the other bands via the sum rules.

On the other hand, it is currently standard to derive localized Wannier functions near the Fermi energy in the first principle's calculation. Therefore, it is necessary to obtain a formula for the orbital magnetic susceptibility expressed in terms of the Wannier functions. In this case, however, we have to be careful to use the truncation. 

\begin{figure}[tb]
 \centering
 \includegraphics[scale=1.0]{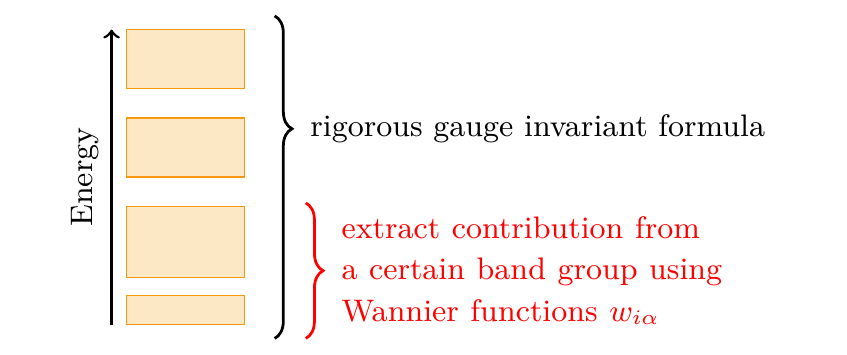}
 \caption{Schematic picture of our scheme. Starting from an $N$ band tight-binding model, we decompose the susceptibility into the contributions from each band group consisting of $M$ ($<N$) bands, and the inter-bandgroup contributions. In this example, $N=4$, and the band group we focus on has two bands ($M=2$).}\label{fig:strategy}
\end{figure}

In this paper, we introduce a scheme to perform truncation of the orbital magnetic susceptibility, or decomposition of the susceptibility into contribution from each band, using localized Wannier functions, motivated from the observation that the correction to the Landau-Peierls susceptibility is from local terms \cite{doi:10.7566/JPSJ.85.064709,doi:10.7566/JPSJ.85.074709}. In order to make the argument as transparent as possible, we work on a multi-orbital tight-binding model instead of a continuum model with periodic potential. Since the number of bands is finite in tight-binding models, we can have exact susceptibility free from truncation errors as a reference. Then, we demonstrate decomposition of the susceptibility for multiband models into contribution from the subset of the bands (see Fig.~\ref{fig:strategy}). We first derive a formula for the susceptibility that includes six terms $\chi_{1}$-$\chi_{6}$ using modified Wannier functions and the Green’s function technique. Importantly, there is an intuitive understanding of this decomposition into six terms: $\chi_{1}$-$\chi_{4}$ are intraband contributions and $\chi_{5}$ and $\chi_{6}$ are interband contributions. From the other point of view, $\chi_1$ is classified as an itinerant contribution, while $\chi_{2}$, $\chi_3$, $\chi_5$, and $\chi_6$ are local contributions, and $\chi_4$ is the cross term between local and itinerant terms. Then, the formula is applied for a couple of simple models. The analysis for the simple models confirms the validity of the formula itself and reveals an interesting feature that the quality of decomposition depends on the degree of localization of the used Wannier functions. 
We will show that, in some cases, the calculated orbital magnetic susceptibility has a 
sizable quantitative error if we apply an inappropriate truncation, or use \textit{not} well-localized Wannier functions. Finally, we clarify the relationship to the Berry curvature in the present scheme of decomposition. 

This paper is organized as follows. First, we derive a formula for the orbital magnetic susceptibility in multiorbital tight-binding models using the Wannier functions, and explain the physical meaning of the terms in the formula. Next, we apply our formula on three models, the honeycomb lattice model with sublattice potential, the honeycomb lattice model with Kekul\'e type distortion, and the decorated square lattice model, and evaluate the validity and usefulness of our formula. Then, the paper is closed by giving discussions and summary.

\section{Derivation of the formula}
\subsection{Model and basic notions}
We start from the tight-binding Hamiltonian
\begin{equation}
  H = \sum_{ij} H_{ij}^{(0)}c^\dagger_ic_j = \sum_{\alpha\beta}H_{0,\alpha\beta}d^\dagger_\alpha d_\beta,
\end{equation}
where $c^\dagger_i$ ($c_i$) is a creation (annihilation) operator for the $i$th orbital, and $d^\dagger_\alpha$ ($d_\alpha$) is a creation (annihilation) operator for the Wannier state indexed by $\alpha$. The lattice points and the spin degrees of freedom are included in $i$ and $\alpha$. Using the Wannier functions $w_{i\alpha}=\langle i|w_{\alpha}\rangle$, $d^\dagger_{\alpha}$ is written as
\begin{equation}
 d^\dagger_\alpha = \sum_ic^\dagger_i w_{i\alpha},
\end{equation}
and the matrices $H^{(0)}$ (in the original basis) and $H_0$ (in the Wannier basis) are related by a unitary transformation as
\begin{equation}
 H_0 = W^\dagger H^{(0)} W,\quad (W)_{i\alpha} = w_{i\alpha}.
\end{equation}
The Wannier functions are constructed so as to have $H_0$ block diagonal in the band group index. Generically, the index $\alpha$ is a combination of the lattice point index and the band group index, and possibly the orbital (or, sublattice) index when the corresponding band group has multiple bands (see Fig.~\ref{fig:strategy}). 

In this paper, we assume that the magnetic field is added through the Peierls substitution in the \textit{original} basis as
\begin{equation}
  H = \sum_{ij} H_{ij}^{(0)}e^{i\frac{q}{\hbar}A_{\bm{r}_i\bm{r}_j}}c^\dagger_ic_j,\label{eq:orig_ham_with_peierls}
\end{equation}
where $q$ is the charge, $\bm{r}_i$ is the position of the $i$th orbital, and $A_{\bm{r}_i\bm{r}_j}$ is defined as 
\begin{equation}
  A_{\bm{r}_i\bm{r}_j} = \int_{C_{\bm{r}_{j}\rightarrow\bm{r}_{i}}} d\bm{r}\cdot \bm{A}(\bm{r}).
\end{equation}
$C_{\bm{r}_j\rightarrow\bm{r}_i}$ is a path from $\bm{r}_j$ to $\bm{r}_i$, which is for simplicity, set to the straight line between $\bm{r}_i$ and $\bm{r}_j$ leading to \cite{PhysRev.84.814}
\begin{equation}
  A_{\bm{r}_i\bm{r}_j} 
   =\int_0^1ds (\bm{r}_i-\bm{r}_j)\cdot\bm{A}(s(\bm{r}_i-\bm{r}_j)+\bm{r}_j).
\end{equation}
Note that the remaining effects of the magnetic field other than the Peierls phase are discussed in Ref.~\citen{doi:10.7566/JPSJ.85.074709}. 

Assuming that the magnetic field is in the $z$-direction, the orbital magnetic susceptibility $\chi$ for free electrons can be obtained by expanding the thermodynamic potential with respect to $B_z$, resulting in the formula \cite{PhysRevB.76.085425,PhysRevB.91.085120}
\begin{equation}
 \chi=k_BT\sum_n\frac{\partial^2F(i\omega_n+\mu,B_z)}{(\partial{B_z})^2},
\end{equation}
where $T$ is temperature, $\omega_n=(2n+1)\pi k_BT$ ($n$: integer) is Matsubara frequency, and $F(z,B_z)$ satisfies
\begin{equation}
 \frac{\partial F(z,B_z)}{\partial z}
  =\mathrm{Tr}G(z,B_z),
\label{eq:free_energy_to_G}
\end{equation}
with $G(z,B_z)$ being the Green function under the magnetic field associated with Hamiltonian Eq.~\eqref{eq:orig_ham_with_peierls}. 

\subsection{Susceptibility in the original basis}
Working with the original basis, it is straightforward to obtain $G(z,B_z)$ in terms of $\check{g}=(z\hat{1}-H^{(0)})^{-1}$ \cite{PhysRevB.76.085425,PhysRevLett.106.045504,PhysRevB.91.085120}. Then, the gauge invariant formula for $\chi$ is written as
\begin{equation}
 \chi =\frac{q^2}{\hbar^2}
  \frac{k_BT}{4}\sum_n \mathrm{Tr}\bigl(
   (\check{\gamma}_x \check{g}\check{\gamma}_y+\check{\gamma}_{xy}) \check{g}\check{\gamma}_x \check{g}\check{\gamma}_y \check{g}
   \bigr)+(x\leftrightarrow y),\label{eq:chi_orig}
\end{equation}
where $\check\gamma_\nu$ is the current operator devided by $q/\hbar$
\begin{equation}
 \check{\gamma}_{\nu}=i[\check{r}_{\nu},H^{(0)}], 
\end{equation}
with
\begin{equation}
 (\check{r}_{\nu})_{ij} = r_{i\nu}\delta_{ij},
\end{equation}
and 
\begin{equation}
\check{\gamma}_{\nu\lambda} = -[\check{r}_{\nu},[\check{r}_{\lambda},H^{(0)}]],
\end{equation}
This is exactly the formula for a tight-binding model with the Peierls phase derived in the previous studies \cite{PhysRevB.76.085425,PhysRevLett.106.045504,PhysRevB.91.085120}.
We will use this formula as a reference to evaluate the usefulness of our decomposition by the formula derived in the following.

\subsection{Susceptibility in the Wannier basis}
Next, we would like to express the Green's function under the magnetic field $G(z,B_z)$ in the Wannier bases, i.e., in terms of $g=(z\hat{1}-H_0)^{-1}$ where $H_0$ is the Wannier Hamiltonian that is block diagonal in the band group index. For this purpose, we find that it is convenient to write the Green function using the Wannier functions as
\begin{equation}
 G_{jk} = \sum_{\alpha\beta}\tilde{w}_{j\alpha}\tilde{G}_{\alpha\beta}\tilde{w}^*_{k\beta},
\end{equation}
where $\tilde{w}_{i\alpha}$ is a Wannier function amended by the Peierls phase\cite{PhysRev.84.814} defined by 
\begin{equation}
 \tilde{w}_{i\alpha}=e^{i\frac{q}{\hbar}A_{\bm{r}_i\bar{\bm{r}}_\alpha}}w_{i\alpha},
\end{equation}
with the Wannier center $\bar{\bm{r}}_{\alpha}$ defined as
\begin{equation}
 \bar{\bm{r}}_{\alpha} = \langle w_\alpha|r|w_\alpha\rangle = \sum_i \bm{r}_iw^*_{i\alpha} w_{i\alpha}. 
\end{equation}
Expanding the Green function in terms of $\tilde{w}_{i\alpha}$ instead of $w_{i\alpha}$ helps us in keeping the gauge invariance. However, in turn, the set $\{\tilde{w}_{i\alpha}\}$ is no longer orthonormalized, and we have to take care of the overlaps between $\tilde{w}_{i\alpha}$. 

Then, some arithmetic gives us (see Appendix for details)
\begin{equation}
 \chi = \frac{q^2}{\hbar^2}k_BT\sum_n(\Xi_{\text{I}} + \Xi_{\text{II}} + \Xi_{\text{III}} + \Xi_{\text{IV}} + \Xi_{\text{V}})
 \label{eq:total_chi}
\end{equation}
with 
\begin{align}
 &\Xi_{\text{I}} = 
  \frac{1}{4}\mathrm{Tr}\bigl(
   (\gamma_x g\gamma_y+\gamma_{xy}) g\gamma_x g\gamma_y g
   \bigr)+(x\leftrightarrow y),\\
 &\Xi_{\text{II}} = -\mathrm{Tr}\Bigl(
  M_z-\frac{i}{2}([\gamma_x,\eta_y]-[\gamma_y,\eta_x])
  \Bigr)g,\\
 &\Xi_{\text{III}} = 
  \mathrm{Tr}
  \Bigl(
  \{S_1^z,L_z\}+\frac{1}{4}[S^z_1,[H_0,S^z_1]]
  \Bigr)g,\\
 &\Xi_{\text{IV}} = -\mathrm{Tr}L_zgL_zg,\\
 &\Xi_{\text{V}} = -i\mathrm{Tr}
  L_zg(\gamma_x g\gamma_y-\gamma_y g\gamma_x)g.
\end{align}
Here
\begin{equation}
 \gamma_{\nu}=i[\hat{r}_{\nu},H_{0}],\quad \gamma_{\nu\lambda} = -[\hat{r}_{\nu},[\hat{r}_{\lambda},H_{0}]],
\end{equation}
with
\begin{equation}
 (\hat{r}_{\nu})_{\alpha\beta} = \bar{r}_{\alpha\nu}\delta_{\alpha\beta},
\end{equation}
and $\bar{r}_{\alpha\nu}$ is the $\nu$ ($=x,y$) component of $\bar{\bm{r}}_\alpha$. Note that in the Wannier basis, $g$, $\gamma_\nu$, and $\gamma_{\nu\lambda}$ are all block diagonal in band groups, which makes it easy to separate intra- and inter-band contributions to $\chi$, as we will do in the following.  
We have also used
\begin{equation}
 L_z=H^z_1-\frac{1}{2}\{S^z_1,H_0\},\quad M_z=H^z_2-\frac{1}{2}\{S^z_2,H_0\},
\end{equation}
\begin{align}
 (S^z_1)_{\alpha\beta} &= i\sum_i\phi^z_{\bar{\bm{r}}_\alpha\bm{r}_{i}\bar{\bm{r}}_\beta}w^*_{i\alpha}w_{i\beta},\label{eq:defS1}\\
 (S^z_2)_{\alpha\beta} &= -\sum_{i}(\phi^z_{\bar{\bm{r}}_\alpha\bm{r}_i\bar{\bm{r}}_\beta})^2w^*_{i\alpha}w_{i\beta},\\
 (H^z_1)_{\alpha\beta} &= i\sum_{ij}\phi^z_{\bar{\bm{r}}_\alpha\bm{r}_i\bm{r}_j\bar{\bm{r}}_\beta}w^*_{i\alpha}H^{(0)}_{ij}w_{j\beta},\\
 (H^z_2)_{\alpha\beta} &= -\sum_{ij}(\phi^z_{\bar{\bm{r}}_\alpha\bm{r}_i\bm{r}_j\bar{\bm{r}}_\beta})^2w^*_{i\alpha}H^{(0)}_{ij}w_{j\beta},\label{eq:defH2}
\end{align}
with
\begin{align}
 \phi^z_{\bar{\bm{r}}_\alpha\bm{r}_{i}\bar{\bm{r}}_\beta}
  &=\frac{1}{2}\bm{e}_z\cdot[(\bar{\bm{r}}_\alpha-\bm{r}_{i})\times(\bar{\bm{r}}_\beta-\bm{r}_{i})],\\
 \phi^z_{\bar{\bm{r}}_\alpha\bm{r}_i\bm{r}_j\bar{\bm{r}}_\beta}
  &=\frac{1}{2}\bm{e}_z\cdot[(\bar{\bm{r}}_\alpha-\bm{r}_j)\times(\bar{\bm{r}}_\beta-\bm{r}_i)],
\end{align}
and
\begin{equation}
 \eta_{\nu} = i[\hat{r}_\nu,S^z_1].
\end{equation}
Note that $S^{z}_{1,2}$ takes account for the finite overlaps of the amended Wannier functions.

Because $H^{(0)}$ and $H_0$ are related to each other by a unitary transformation and the formula involves the trace, one may think that the formula should look the same in any basis. Actually, the contribution $\Xi_{\text{I}}$ shares the same form with Eq.~\eqref{eq:chi_orig}. However, we have to note that 
\begin{equation}
 \hat{r}_{\nu} \neq W^\dagger \check{r}_{\nu} W\label{eq:rvsr}
\end{equation}
holds since $\hat{r}_{\nu}$ only picks up the diagonal elements by definition, while $(W^\dagger \check{r}_{\nu} W)_{\alpha\beta} = \sum_ir_{i\nu}w^*_{i\alpha}w_{i\beta}$ is generically finite for $\alpha\neq\beta$. This is the reason why the additional terms appear in Eq.~\eqref{eq:total_chi}. This means that $\Xi_{\text{II}}$, $\Xi_{\text{III}}$, $\Xi_{\text{IV}}$, and $\Xi_{\text{V}}$ are the correction terms that reflect $\hat{r}_{\nu} \neq W^\dagger \check{r}_{\nu} W$. We will come back to this point later. 

\subsection{Intra- and interband contributions}
Now, we decompose $\chi$ into various kinds of contributions, such as intra- and inter-band contributions. For $\Xi_{\text{I}}$, since $g$, $\gamma_{\nu}$, and $\gamma_{\nu\lambda}$ are diagonal in band groups, we can simply extract the contribution from the band group $i$ by introducing
\begin{equation}
 \Xi_1^{[i]} = \frac{1}{4} \mathrm{Tr}^{[i]}\bigl(
   (\gamma_x g\gamma_y+\gamma_{xy}) g\gamma_x g\gamma_y g
   \bigr)+(x\leftrightarrow y),
\end{equation}
where $\mathrm{Tr}^{[i]}$ denotes the partial trace over the band group $i$. $\Xi_{\text{II}}$ and $\Xi_{\text{V}}$ also have only intraband contributions because of the trace structure, although $M_z$, $L_z$, and $\eta_{\nu}$ can have matrix elements between different band groups. 

On the other hand, $\Xi_{\text{III}}$ and $\Xi_{\text{IV}}$ have both intra- and interband contributions, since they are in the second order with respect to $L_z$ or $S_1^z$ (or their cross terms). Note that it is sufficient to think of pairs of band groups due to the trace structure. Namely, starting from a band group $i$, and going to another band group $j$ by $L_z$ or $S_1^z$, then it has to be back on the band group $i$ by the second $L_z$ or $S_1^z$ to complete the trace. Therefore, for the intraband contribution of $\Xi_{\text{II}}+\Xi_{\text{III}}$, we introduce
\begin{multline}
 \Xi_2^{[i]} = 
 -\mathrm{Tr}^{[i]}\Bigl(
  M_z-\frac{i}{2}([\gamma_x,\eta_y]-[\gamma_y,\eta_x])
  \Bigr)g\\
  +
  \mathrm{Tr}^{[i]}
  \Bigl(
  \{S_1^{z[i]},L_z^{[i]}\}+\frac{1}{4}[S^{z[i]}_1,[H_0,S^{z[i]}_1]]
  \Bigr)g,\label{eq:Xi2}
\end{multline}
where $X^{[i]}$ means a matrix constructed from $X$ by leaving only the matrix elements within the band group $i$. Similarly, for the intra band contributions of $\Xi_{\text{IV}}$ and $\Xi_{\text{V}}$, we introduce
\begin{equation}
 \Xi_3^{[i]} = -\mathrm{Tr}^{[i]}L_z^{[i]}gL_z^{[i]}g,
\end{equation}
and
\begin{equation}
 \Xi_4^{[i]} = -i\mathrm{Tr}^{[i]}
  L_zg(\gamma_x g\gamma_y-\gamma_y g\gamma_x)g.
\end{equation}
Finally, for the interband contributions of $\Xi_{\text{III}}$ and $\Xi_{\text{IV}}$, we introduce
\begin{equation}
 \Xi_5^{[i:j]} = 
  \mathrm{Tr}
  \Bigl(
  \{S_1^{z[i:j]},L_z^{[i:j]}\}+\frac{1}{4}[S^{z[i:j]}_1,[H_0,S^{z[i:j]}_1]]
  \Bigr)g,
\end{equation}
and
\begin{equation}
 \Xi_6^{[i:j]} = -\mathrm{Tr}L_z^{[i:j]}gL_z^{[i:j]}g,
\end{equation}
where $X^{[i:j]}$ denotes a matrix constructed from $X$ by leaving only the matrix elements connecting the band groups $i$ and $j$. 

Now, we decompose $\chi$ as 
\begin{equation}
 \chi = \chi_1+\chi_2+\chi_3+\chi_4+\chi_5+\chi_6,\label{eq:chi_1to6}
\end{equation}
where 
\begin{equation}
 \chi_a = k_B T\sum_n\sum_i\Xi_a^{[i]}\label{eq:karnel_intra}
\end{equation}
for $a=$\{1,2,3,4\}, and 
\begin{equation}
 \chi_a = k_B T\sum_n\sum_{i<j}\Xi_a^{[i:j]}\label{eq:karnel_inter}
\end{equation}
for $a=$\{5,6\}. For later use, we also define the susceptibility contributed from a set of band groups $X$ as
\begin{equation}
 \chi^{\{X\}} = k_B T\sum_n
\Bigl(
  \sum_{i\in X}\sum_{a=1}^4 \Xi_a^{[i]}+
  \sum_{i,j\in X,i<j}\sum_{a=5}^6 \Xi_a^{[i:j]}
 \Bigr).\label{eq:chi_group}
\end{equation}

\subsection{Physical interpretation of each contribution}
\begin{table}[tb]
 \centering
 \caption{Decomposition of the susceptibility. ``Intra'' and ``inter'' in the table denote ``intra-bandgroup'' and ``inter-bandgroup'', respectively. ``Cross term'' means the cross term between itinerant motion and local moment. The column ``in-gap'' shows whether each contribution can be finite when the chemical potential is in the energy gap \textit{between the band groups} at zero temperature. Note that all the terms can be finite in the energy gap which may exist \textit{inside a band group}. (See Fig.~\ref{fig:strategy}.)}\label{tab:interpretation}
 \begin{tabular}{ccccc}
  \hline
  & band & type & origin & in-gap \\
\hline
  $\chi_1$ & intra & itinerant & electron hopping &  \\
  $\chi_2$ & intra & local & atomic diamagnetism & \checkmark \\
  $\chi_3$ & intra & local & orbital Zeeman \& van Vleck & \\
  $\chi_4$ & intra & both & cross term & \\
  $\chi_5$ & inter & local & atomic diamagnetism & \checkmark \\
  $\chi_6$ & inter & local & van Vleck & \checkmark \\
  \hline
 \end{tabular}
\end{table}
Here, we discuss the physical meaning of the decomposition. As we have already seen in the derivation, $\chi_{1}$-$\chi_{4}$ are the intraband contributions, while $\chi_{5}$ and $\chi_{6}$ are the interband contributions. 

In particular, $\chi_1$ has the same form as Eq.~\eqref{eq:chi_orig}, and therefore, is interpreted as a contirbution from the Peierls substitution in the Wannier basis. Since the Peierls phase affects the hopping, $\chi_1$ is, more or less, a response from electrons hopping around, and thus, we regard $\chi_1$ as itinerant contribution. 

On the other hand, $\chi_2$, $\chi_3$, $\chi_5$, and $\chi_6$ are finite in the presence of $H^z_{1,2}$ and $S^z_{1,2}$. By definition, $H^z_{1,2}$ and $S^z_{1,2}$ are short ranged if the Wannier functions are well localized. Therefore, we categorize these terms as local contributions. When we look at each term more closely, we can see that $\chi_2$ and $\chi_5$ include a single $g$ in the formula, and give contributions proportional to $f(E_\alpha)$, where $f$ is the Fermi distribution function and $E_\alpha$ is eigenenergy in the band group. Namely, these terms depend on the filling of the band group, and can be interpreted as atomic diamagnetism generalized to the Wannier orbitals. In particular, $\chi_5$ renormalizes atomic diamagnetism via interband effects. Here, we use the term ``diamagnetism'' to match it to the conventional terminology in the case of real atoms, but in the case of the Wannier functions in tight-binding models, the sign of this term is not necessarily negative, and there can be ``atomic paramagnetism''. Especially for a tight-binding model where the energy range of the bands is bounded, the orbital magnetic susceptibility induced by the Peierls phase has to vanish at zero temperature when the chemical potential is higher than the upper limit of the band energy. This is to satisfy the susceptibility sum rule \cite{PhysRevLett.112.026402}. Therefore, the atomic diamagnetism associated with some band has to be compensated by the atomic ``paramagnetism'' associated with the other band.

In contrast to $\chi_2$ and $\chi_5$, $\chi_3$ and $\chi_6$ include two $g$'s in the formula, and give contributions proportional to $(f(E_\alpha)-f(E_\beta))/(E_\alpha-E_\beta)$. These terms can be interpreted as the contributions from the fluctuation of the local moment $L_z$, which account orbital Zeeman type susceptibility and van Vleck type susceptibility. Note that each band group can have multiple bands in our formalism, and thus, $\chi_3$, categorized into intraband component, can also contain van Vleck type responses between the bands inside the band group. 

Finally, interpreting $L_z$ as a local moment, $\chi_4$ can be seen as a cross coupling between the local moment and itinerant motion of electrons by hopping, which is similar to the cross coupling between spins and itinerant motion that arises in models with spin Zeeman term \cite{PhysRevResearch.3.013058}.

These interpretations are summarized in Table~\ref{tab:interpretation}. It is worth noting that among the intraband terms $\chi_{1,2,3,4}$, only $\chi_2$ can have finite value in the zero temperature limit when the chemical potential is outside of the energy range of the corresponding band, since $\chi_2$ depends on the orbital filling itself rather than some fluctuations. Therefore, $\chi$ at the band gap should be from $\chi_{2,5,6}$ at zero temperature.

\section{Application to Tight-Binding Models}
Now, we move on to some demonstrations of the derived formula applied to some tight-binding models. The Wannier functions play a central role in this study. In the following, the Wannier functions are derived by projecting candidate wave functions $w^{(c)}_{il}$ to the Hilbert space spanned by a specific band group. In $w^{(c)}_{il}$, $i$ denotes the $i$th site just as $i$ in $w_{i\alpha}$ above, while $l$ is for indexing the candidates. This procedure is often used in deriving an initial guess for a Wannier function in the well known method to obtain maximally localized Wannier functions \cite{PhysRevB.56.12847}. Here, we adapt a method to derive an initial guess for the tight-binding models. First, we compute Bloch wave functions $\psi_{i,n\bm{k}}$ for the target band group with the momenta $\bm{k}$ on a regular grid in the Brillouin zone, where $n$ specifies a band in the band group. Then, we derive a matrix $A_{\bm{k}}$ whose matrix elements are $(A_{\bm{k}})_{nl}=\sum_i\psi^*_{i,n\bm{k}}w^{(c)}_{il}$, and perform its singular value decomposition as $A_{\bm{k}}=U_{\bm{k}}\Lambda_{\bm{k}}V^{{\dagger}}_{\bm{k}}$. Using these unitary matrices $U_{\bm{k}}$ and $V_{\bm{k}}$, the initial guess Wannier functions can be fixed as
\begin{equation}
w_{i,\underline{\alpha}\bm{R}}=\sum_{\bm{k}}\sum_n e^{-i\bm{k}\cdot\bm{R}}\psi_{i,n\bm{k}}(U_{\bm{k}}V^\dagger_{\bm{k}})_{n\underline{\alpha}},
\end{equation}
with appropriate normalization, where the lattice point dependence of the Wannier function is explicitly indicated by replacing $\alpha\rightarrow\underline{\alpha}\bm{R}$ with $\underline{\alpha}$ for the degrees of freedom other than the lattice points. The unitarity of $U_{\bm{k}}V^\dagger_{\bm{k}}$ combined with the orthogonality of the Bloch wave functions ensures the orthogonality of the obtained Wannier functions.
In the following, we only work with simple models, and this ``initial guess'' already shows fairly good localization as we will see shortly. 

\subsection{Honeycomb lattice with sublattice potential}
\begin{figure}[tb]
 \centering
 \includegraphics[scale=1.0]{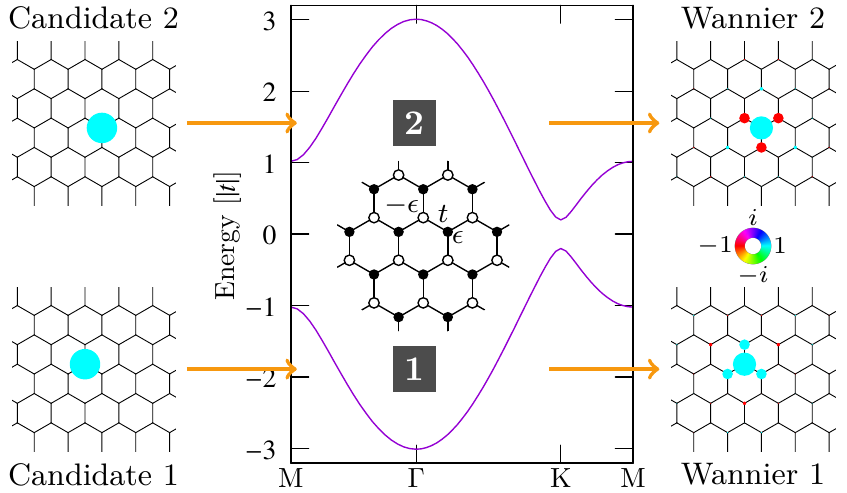}
\caption{Band structure, candidate wave functions, and Wannier functions for the honeycomb lattice model with a sublattice dependent potential. (Inset in the middle shows $H^{(0)}_{ij}$ in the original basis.) Finite site potential gaps out the Dirac cone. Candidate 1 and 2 on the left are initial guess to obtain the Wannier functions. Wanner 1 and 2 on the right represent the Wannier functions for the lower and the upper bands, respectively.}\label{fig:typeAwan}
\end{figure}
The first model we tackle is a honeycomb lattice model with a sublattice dependent potential \cite{PhysRevLett.53.2449}. As illustrated in the inset in the middle of Fig.~\ref{fig:typeAwan}, the matrix $H^{(0)}_{ij}$ in the original basis consists of the nearest neighbor pairs of sites $t$, and A (B) sublattice potential $+\epsilon$ ($-\epsilon$) for $i=j$. The finite $\epsilon$ induces a gap at the Dirac cones for the pristine honeycomb lattice model, enabling us to decompose the bands into the lower and the upper bands. We use $t=-1$ and $\epsilon=0.2$, resulting in the gap size of $0.4|t|$. Because the sublattice potential accounts for the gap, the candidate wave function is chosen to be completely localized on a single A (or, B) sublattice. For instance, in order to have a Wannier function for the lower band, we set $w^{(c)}_{il}=1$ only for the B sublattice (whose site potential is $-\epsilon$ with $\epsilon>0$) in the unit cell at the origin ($\bm{R}=0$) and $w^{(c)}_{il}=0$ for the rest of the sites. We can see from Fig.~\ref{fig:typeAwan} that the obtained Wannier functions are fairly well localized.

\begin{figure}[tb]
 \centering
 \includegraphics[scale=1.0]{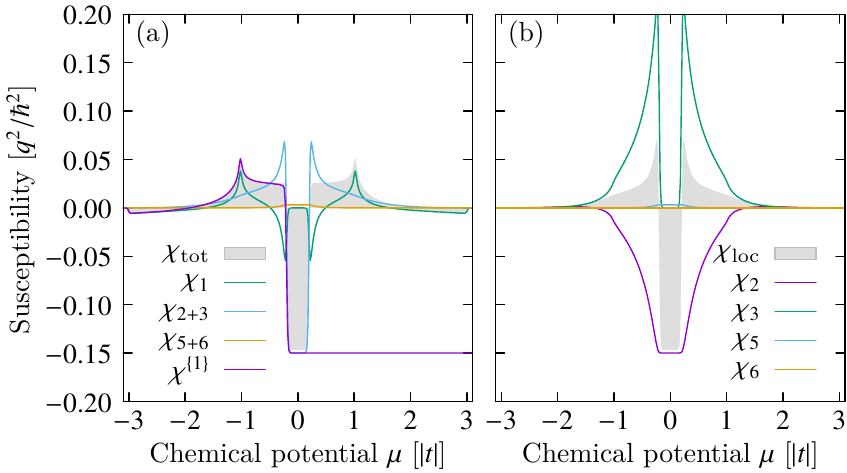}
\caption{(a) Susceptibility at $T=0.01|t|$ as a function of chemical potential for the honeycomb lattice model with a sublattice dependent potential. $\chi_{2+3}$ and $\chi_{5+6}$ are abbreviations of $\chi_2+\chi_3$ and $\chi_5+\chi_6$, respectively. Gray shadow of (a) shows the exact $\chi$ obtained by Eq.~\eqref{eq:chi_orig}. (b) Chemical potential dependence of $\chi_2$, $\chi_3$, $\chi_5$, and $\chi_6$. Their total is shown as $\chi_{\text{loc}} \equiv \chi_{2+3}+\chi_{5+6}$.
}\label{fig:typeAchi}
\end{figure}
Figure~\ref{fig:typeAchi} summarizes the calculated susceptibility $\chi$ at $T=0.01|t|$ as a function of the chemical potential $\mu$. As a reference, exact $\chi$ obtained with Eq.~\eqref{eq:chi_orig} in the original basis is shown as a gray shadow in Fig.~\ref{fig:typeAchi}(a). It is confirmed that the total $\chi$ obtained with Eqs.~\eqref{eq:chi_1to6}-\eqref{eq:karnel_inter} matches with the exact result within the line width in the scale of Fig.~\ref{fig:typeAchi}(a). (Not shown to avoid making the figure busy.) 

A notable feature in Fig.~\ref{fig:typeAchi}(a) is that $\chi^{\{1\}}$, which is the contribution from the band group 1 (lower band), nicely approximates the exact result for $\mu<0$. The same feature is also noticed by the smallness of the interband contribution $\chi_5+\chi_6$ over the whole range of $\mu$. This means that the decomposition into the contribution from each band is successful in this model with our choice of the Wannier functions.

Figures~\ref{fig:typeAchi}(a) and \ref{fig:typeAchi}(b) also tell us that the diamagnetic responce at the gap ($\mu\sim 0$) is from $\chi_2$, intraband contribution corresponding to the atomic diamagnetism of the Wannier function. Among the terms in Eq.~\eqref{eq:Xi2} relevant to $\chi_2$, the terms involving $([\gamma_x,\eta_y]-[\gamma_y,\eta_x])$ and $[S^{z[i]}_1,[H_0,S^{z[i]}_1]]$ are zero in this specific model, and $\chi_2$ is contributed by the terms involving $M_z$ and $\{S_1^{z[i]},L_z^{[i]}\}$. As we have noted, the atomic diamagnetism associated with the lower band is compensated by the atomic paramagnetism associated with the upper band, and this is the reason that $\chi_2$ has positive slope when $\mu$ is in the energy range of the upper band. The same kind of compensation is found in the following examples as well.

\subsection{Honeycomb lattice with Kekul\'e type modulation}
\begin{figure*}[tb]
 \centering
 \includegraphics[scale=1.0]{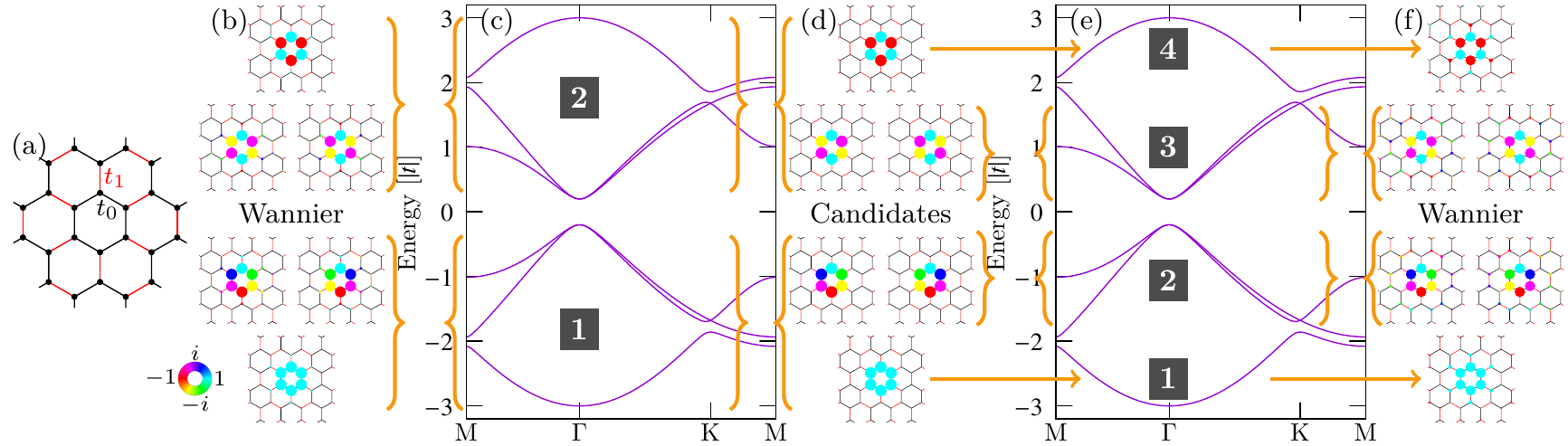}
\caption{Band structure, candidate wave functions, and Wannier functions for the honeycomb lattice model with Kekul\'e type distortion. (a) Schematic picture of the model. (b,f) The obtained Wannier functions for the case (b) with the two band groups and (f) with the four band groups, respectively. (c,e) Band structure with band groups marked with the numbers. (d) Candidates are initial guess to obtain the Wannier functions.}\label{fig:typeCwan}
\end{figure*}
The next model is a honeycomb lattice model with Kekul\'e type hopping texture \cite{PhysRevLett.98.186809,Wu:2016aa,Kariyado:2017aa}. Specifically, we introduce two types of hopping, $t_0$ and $t_1$, as illustrated in Fig.~\ref{fig:typeCwan}(a). With this texture, the six site cluster connected by $t_0$ bonds forms a unit cell, resulting in the Brillouin zone folding and gapping out the Dirac cones in the pristine honeycomb lattice. In this study, we set $t_0=t-\delta$, $t_1=t+2\delta$, $t=-1$, and $\delta=1/15$, which gives a gap of $0.4|t|$ at zero energy. The state with $\delta=1/15$ is adiabatically connected to the state with $t_1\rightarrow 0$ ($t_0\neq 0$). In this limit, the system consists of six site clusters decoupled with each other. Then, each cluster can be regarded as a one-dimensional periodic chain with six sites, whose eigenstates can be expressed as plane waves along the chain $\psi^{\text{cluster}}_n\sim e^{ik_nx}$ with $k_n=\frac{2\pi}{a}\frac{n}{6}$, where $a$ denotes the unit length and $n=\{0,\pm 1,\pm 2,3\}$. 

\begin{figure}[tb]
 \centering
 \includegraphics[scale=1.0]{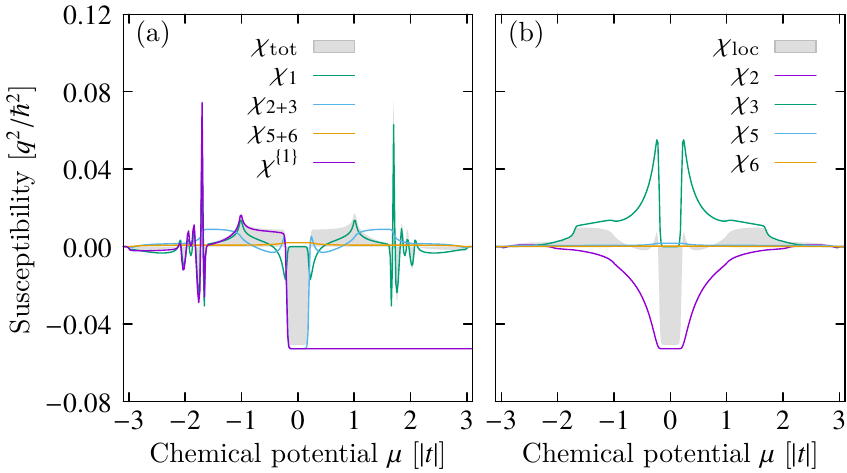}
\caption{Susceptibility at $T=0.01|t|$ as a function of chemical potential for the honeycomb lattice model with Kekul\'e type distortion, where the six bands are decomposed into two band groups. 
}\label{fig:typeCchi}
\end{figure}
We use $\psi^{\text{cluster}}_n$ with $n=0,\pm 1$ as the candidate wave functions to build Wannier functions for the three bands at $E<0$. [See Figs.~\ref{fig:typeCwan}(b)-\ref{fig:typeCwan}(d)] The obtained Wannier functions look fairly well localized [Fig.~\ref{fig:typeCwan}(b)]. Then, these Wannier functions are used to decompose the susceptibility in the contributions from the band group 1 (bands at $E<0$) and the band group 2 (bands at $E>0$), which are shown in Fig.~\ref{fig:typeCchi}. As in the previous example, the exact $\chi$ obtained with the original basis is shown as gray shade in Fig.~\ref{fig:typeCchi}(a). Similarly to the previous example, $\chi^{\{1\}}$ approximates the exact result very well for $\mu<0$ [Fig.~\ref{fig:typeCchi}(a)], and the diamagnetism in the gap ($\mu\sim 0$) is mostly contributed from $\chi_2$, signaling that the decomposition is effective and useful in this model with our choice of the Wannier functions.

Closely looking at the band structure, we notice that there is a small gap between the lowest band and the second lowest band [Fig.~\ref{fig:typeCwan}(e)]. Therefore, in principles, it is possible to decompose the bands into four band groups, the lowest, the second lowest (having two bands), the second highest (having two bands), and the highest band groups, instead of the two band groups ($E<0$ and $E>0$). In the decomposition into the two groups, the condition for the Wannier function is that three Wannier functions should cover the space spanned by the three bands. On the other hand, in the decomposition into the four groups, the condition is that one Wannier function should strictly generate the space spanned by the lowest band, and two Wannier functions should cover the space spanned by the second lowest band group. Specifically, we use $\psi^{\text{cluster}}_0$ for the lowest band, and $\psi^{\text{cluster}}_{\pm 1}$ for the second lowest band group. That is, there is stronger restriction in the four-group decomposition than the two-group decomposition. Because of this stronger restriction, the Wannier functions for the four-group decomposition [Fig.~\ref{fig:typeCwan}(f)] are less localized compared with those for the two-group decomposition [Fig.~\ref{fig:typeCwan}(b)]. 

\begin{figure}[tb]
 \centering
 \includegraphics[scale=1.0]{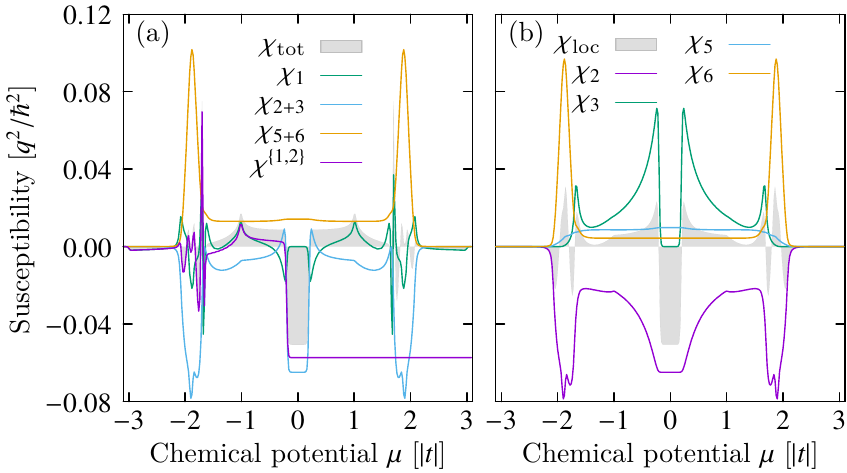}
\caption{Susceptibility at $T=0.01|t|$ as a function of chemical potential for the honeycomb lattice model with Kekul\'e type distortion, where the six bands are decomposed into four band groups.}\label{fig:typeDchi}
\end{figure}
The susceptibility obtained with the less localized Wannier functions is summarized in Fig.~\ref{fig:typeDchi}. As a sanity check, it is confirmed that we still have good agreement with the exact result if we collect all the terms, which strongly supports the validity of our formulation itself. However, when it comes to the quality of the approximation, the quality of the Wannier functions does matter. If we focus on $\chi^{\{1,2\}}$ in Fig.~\ref{fig:typeDchi}(a), which corresponds to the contribution from the band at $E<0$ (including the interband contribution between the band group 1 and 2), we can see some deviations from the exact result for $E<0$. Inside the gap ($\mu\sim 0$), the deviation is from both $\chi_5$ and $\chi_6$ [Fig.~\ref{fig:typeDchi}(b)]. Another notable feature is found at $\mu\sim -2$, where there is a small gap between the lowest and the second lowest band. We find cancellation between the diamagnetic contribution from the intraband term $\chi_2$ and the paramagnetic contribution from the interband term $\chi_6$ [Figs.~\ref{fig:typeDchi}(a) and \ref{fig:typeDchi}(b)]. Because of the small gap, the van Vleck type response is enhanced, but we know from the exact result by the original basis that there is no paramagnetic peak, and this discrepancy is resolved by the intraband atomic diamagnetism of the Wannier functions. 

In short, this example tells us that the Wannier functions have to be carefully chosen to make the decomposition into each band effective. Both of $\chi^{\{1\}}$ in Fig.~\ref{fig:typeCchi} and $\chi^{\{1,2\}}$ in Fig.~\ref{fig:typeDchi} are for the contribution from the bands in $E<0$. [See Figs.~\ref{fig:typeCwan}(c) and \ref{fig:typeCwan}(e) for the numbering of the bands.] However, looking at the susceptibility at $\mu=0$, $\chi^{\{1\}}$ in Fig.~\ref{fig:typeCchi} is much closer to the exact result shown by the gray shadow than $\chi^{\{1,2\}}$ in Fig.~\ref{fig:typeDchi}, and this difference is originated from the difference in the Wannier functions [Figs.~\ref{fig:typeCwan}(b) and \ref{fig:typeCchi}(f)].

\subsection{Decorated square lattice model}
\begin{figure}[tb]
 \centering
 \includegraphics[scale=1.0]{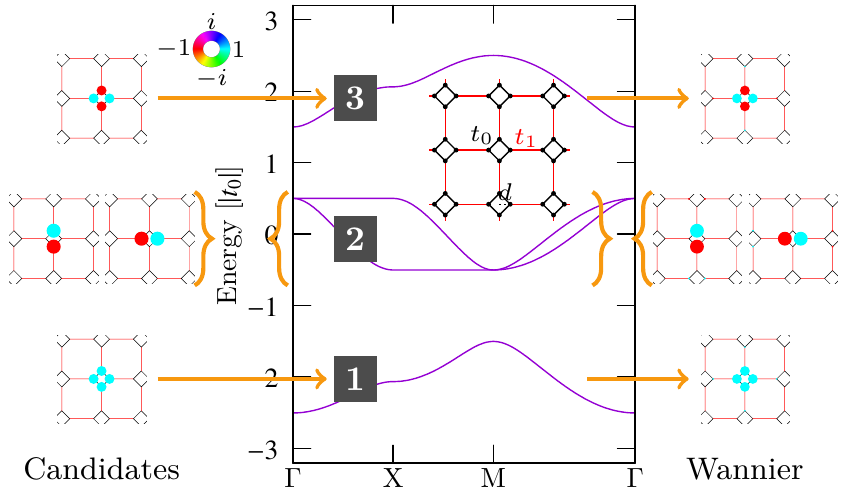}
\caption{Band structure, candidate wave functions, and Wannier functions for the decorated square lattice model. Candidates on the left are initial guess to obtain the Wannier functions, while the obtained Wannier functions are shown on the right. The inset in the middle shows the schematic  description of the model.}\label{fig:typeEwan}
\end{figure}

Before going to the discussion part, we work on a lattice other than the honeycomb type. The model we handle here is a decorated square lattice model \cite{PhysRevB.82.085106,PhysRevB.90.085132}, which is a square network of diamond-shape four-site clusters as illustrated in the inset of Fig.~\ref{fig:typeEwan}. We have three parameters, $t_0$, $t_1$, and $d$. $t_0$ and $t_1$ are the hopping parameters within the four-site cluster and between the clusters, respectively, while $d$ determines the size of the diamond-shape cluster, i.e., the distance between the center and the vertex of the diamond. We set $t_0=-1$, $t_1=-0.5$, and $d=1/5$ (in the unit of the lattice constant). With this choice, the bands are decomposed into three groups. It is noticed that three bands are dominated by $s$-, $p_{x,y}$-, $d_{x^2-y^2}$-like orbitals on the four site cluster, and thus, we use candidate wave functions completely localized on a single cluster with $s$-, $p_{x,y}$-, $d_{x^2-y^2}$-like symmetry to obtain the Wannier functions (Fig.~\ref{fig:typeEwan}). 

\begin{figure}[tb]
 \centering
 \includegraphics[scale=1.0]{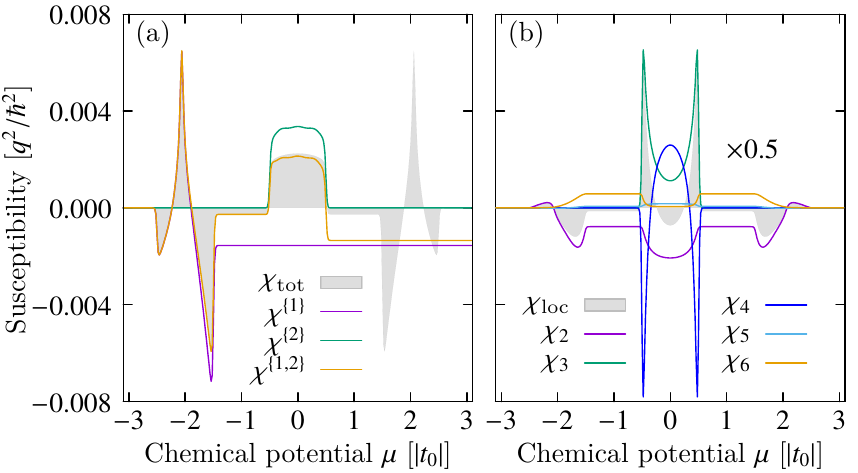}
 \caption{Susceptibility at $T=0.01|t_0|$ as a function of chemical potential for the decorated square lattice model.}
 \label{fig:typeEchi}
\end{figure}
The calculated susceptibility is summarized in Fig.~\ref{fig:typeEchi}. As the previous examples, the exact result is shown as gray shade in Fig.~\ref{fig:typeEchi}(a). We notice that $\chi^{\{1\}}$, which is the contribution from the band group 1 (the lowest band), gives a good approximation for the lower half of the lowest band ($\mu<-2$), and $\chi^{\{1,2\}}$, which represents the contributions from the band groups 1 and 2 (including the interband terms between the band groups 1 and 2), nicely agrees with the exact result for $\mu<0$. However, $\chi^{\{1\}}$ fails to reproduce the exact result in the gap between the band groups 1 and 2, and the deviation is from $\chi_6$ [Fig.~\ref{fig:typeEchi}(b)]. Another interesting feature found in Fig.~\ref{fig:typeEchi}(b) is that $\chi_4$, which is the cross term between the itinerant motion and the local moment, compensates the sharp peaks in $\chi_3$ to restore the exact result, telling us that the cross term cannot be simply neglected. 

In Fig. 8(a), we focus on the region $-1.5<\mu<-0.5$ where the chemical potential is 
located inside the gap between the lowest and the second-lowest bands. 
In this region, $\chi^{\{ 1 \} }$ has a negative finite value while $\chi^{\{ 2 \} }$ is zero. 
However, $\chi^{\{1\}}$ shows a significant deviation from the correct susceptibility given by the gray shadow. This means that $\chi^{\{1\}}+\chi^{\{2\}}$ is not a good approximation for the susceptibility in this region of $\mu$. On the other hand, $\chi^{\{1,2\}}$ is a good approximation in the same region. As we can see from Eq.~\eqref{eq:chi_group}, $\chi^{\{1,2\}}$ differs from $\chi^{\{1\}}+\chi^{\{2\}}$ by
\begin{equation}
    k_BT\sum_n\bigl(\Xi_5^{[1:2]}+\Xi_6^{[1:2]}\bigr),
\end{equation}
namely by the interband contributions. This is a typical example in which the decomposition method has a crucial effect on the calculated susceptibility.

\section{Discussion and Summary}
\subsection{Wannierization dependence}
Along the course of this study, we learned that the quality of the Wannier functions has no effect on the total susceptibility, while it strongly affects the quality of the decomposition. Naively, one may find that this is a condradiction, since generically two sets of Wannier functions are related by a unitary transformation and our formula involves matrix traces. The origin of the Wannier function dependence of the decomposition can be traced back to the definitions of $S^z_{1,2}$ and $H^z_{1,2}$, i.e., Eqs.~\eqref{eq:defS1}-\eqref{eq:defH2}. Since $\phi^z_{\bar{\bm{r}}_\alpha\bm{r}_i\bar{\bm{r}}_\beta}$ and $\phi^z_{\bar{\bm{r}}_\alpha\bm{r}_i\bm{r}_j\bar{\bm{r}}_\beta}$ depends on the Wannier centers $\bar{\bm{r}}_{\alpha,\beta}$, which are computed from the given Wannier functions, Eqs.~\eqref{eq:defS1}-\eqref{eq:defH2} are acutally not simple unitary transformation from the original basis. Then, $S^z_{1,2}$ and $H^z_{1,2}$ for two different sets of Wannier functions may not be related by a unitary transformation, even if the two sets of Wannier functions span the same band groups. 

The Wannierization dependence gives us a chance to optimize Wannier functions for susceptibility decomposition. Generically speaking, the decomposition becomes particularly powerful if we can suppress the interband contribution. Looking at the definitions, we notice that one of the interband contribution $\chi_5$ vanishes if $S^z_1=0$. Note that the matrix element $(S^z_1)_{\alpha\beta}$ for pairs of Wannier functions sharing the Wannier center is zero since $\phi^z_{\bar{\bm{r}}_\alpha\bm{r}_i\bar{\bm{r}}_\beta}=0$ when $\bar{\bm{r}}_\alpha=\bar{\bm{r}}_\beta$. Then, if the given Wannier functions are very well localized such that the overlap between the pairs of Wannier functions \textit{not} sharing the Wannier center is small, we expect that $\chi_5$ is small due to the small $S^z_1$. In the case where $\chi_5$ is negligible, we have $\chi=\chi_2+\chi_6$ at zero temperature when $\mu$ is in some energy gap. Using the hermitian nature of $L^z$, it is straightforward to show that the remaining interband contribution $\chi_6$ is always positive. Then, minimizing the interband contribution $\chi_6$ means minimizing the negative, or diamagnetic contribution to $\chi_2$ since the sum is fixed. Very naively, we can make the diamagnetic contribution in $\chi_2$ small by using well localized Wannier functions, since $\chi_2$ corresponds to the atomic diamagnetism of Wannier orbitals. From all above, we speculate that using compact Wannier functions results in better decomposition with smaller interband contributions. However, we have to do much more extensive surveys to be conclusive. 

\subsection{Berry connections}
It has been briefly noted that Eq.~\eqref{eq:rvsr} is a key to obtain our formula. In order to elaborate this argument, we introduce $\hat{q}_\mu$ as
\begin{equation}
 W^\dagger \check{r}_\mu W = \hat{r}_\mu +\hat{q}_\mu. \label{eq:def_q}
\end{equation}
By definition, we have $(\hat{q}_\mu)_{\alpha\beta}=\sum_ir_{i\mu}w^*_{i\alpha} w_{i\beta}$ ($\alpha\neq\beta$), i.e., $\hat{q}_\mu$ is for the offdiagonal matrix elements of the position operator in the Wannier basis. When the system has lattice translation symmetry, it is convenient to include lattice point positions in the index to specify the Wannier functions as $w_{i\alpha}\rightarrow w_{i,\underline{\alpha}\bm{R}}$, where $\bm{R}$ and $\underline{\alpha}$ denote the lattice point and the remaining degrees of freedom in the unit cell, respectively. Then, we can construct a Bloch wave function corresponding to $w_{i,\underline{\alpha}\bm{R}}$ as $\psi_{i,\underline{\alpha}\bm{k}}=\sum_{\bm{R}}e^{i\bm{k}\cdot\bm{R}}w_{i,\underline{\alpha}\bm{R}}$. Then, it is known\cite{BLOUNT1962305,PhysRevB.56.12847} that we have
\begin{equation}
 (\hat{q}_\mu)_{\underline{\alpha}\bm{R}\underline{\beta}\bm{R}'}
  = \frac{i}{N}\sum_{\bm{k}}e^{i\bm{k}\cdot(\bm{R}-\bm{R}')}\sum_iu^*_{i,\underline{\alpha}\bm{k}}\frac{\partial u_{i,\underline{\beta}\bm{k}}}{\partial k_\mu},\label{eq:Berry}
\end{equation}
where $u_{i,\underline{\alpha}\bm{k}}$ is the periodic part of the Bloch wave function extracted as $u_{i,\underline{\alpha}\bm{k}}=e^{-i\bm{k}\cdot\bm{r}_i}\psi_{i,\underline{\alpha}\bm{k}}$. Equation~\eqref{eq:def_q} also gives
\begin{equation}
 W^\dagger \check{\gamma}_\mu W = \hat{\gamma}_\mu +i[\hat{q}_\mu,H_0].\label{eq:gvsg} 
\end{equation}
Comparing Eq.~\eqref{eq:chi_orig} and Eqs.~\eqref{eq:chi_1to6}-\eqref{eq:karnel_inter}, we know that $\chi_{2}$-$\chi_{6}$ contains at least one $\hat{q}_\mu$, namely, the quantity $\hat{q}_\mu$ that accounts for the difference between $W^\dagger\check{r}_\mu W$ and $\hat{r}_\mu$ indeed generates $\chi_{2}$-$\chi_{6}$. Apparently from Eq.~\eqref{eq:Berry}, $\hat{q}_\mu$ is related to the Berry connection, and it is interesting that the terms $\chi_{2}$-$\chi_{6}$ are generated because of that. Note that the second term at the right hand side of Eq.~\eqref{eq:gvsg} corresponds to $p_{\ell\ell',\mu}$ in Ref.~\citen{doi:10.7566/JPSJ.84.124708}, which plays an essential role in applying the sum rule. Also, it is worth noting that the Berry curvature plays an important role in the semiclassical theory of the orbital magnetic susceptibility \cite{PhysRevB.91.214405} through the anomalous velocity. These observations suggest that our formula complements the previous formulae relying on the Berry connection. Equipped with our formalism, we can use the matrices $S^z_{1,2}$ and $H^z_{1,2}$ to calculate the susceptibility, instead of the Bloch wave functions and their connections. Which formula, the formula in this paper or the former ones, does perform better may depend on situations. However, the current formula gives intuitive understanding of the origin of the susceptibility as summarized in Table~\ref{tab:interpretation}.

\subsection{Summary and Outlook}
To summarize, we have proposed a new formula to decompose the orbital magnetic susceptibility in the contributions from each band, and we have demonstrated the decomposition in some simple models. Here, we took a tight-binding model as a starting point. Since the size of the Hilbert space of the tight-binding model is finite, we can always get a rigorous result without any concerns about truncation error, making the evaluation of the decomposition easy and clear. We believe that almost the same procedure can be applicable to continuum models. Namely, once the explicit formula for $S^z_{1,2}$ and $H^z_{1,2}$ are obtained, the rest of the derivation will not be changed. Nowadays, it becomes more and more common to use Wannier functions obtained in the first-principles calculation for analyzing properties of real materials. In this situation, it is an interesting and important future work to demonstrate the decomposition in continuum models, having a generic formula working in real materials as an ultimate goal. 

\begin{acknowledgments}
The work was partially supported by JSPS KAKENHI Grants No.~JP17K14358 (T.K.), No.~JP20K03844 (T.K.), and No.~JP18H01162. Part of the computations in this work has been done using the facilities of the Supercomputer Center, the Institute for Solid State Physics, the University of Tokyo.
\end{acknowledgments}

\appendix

\section{Derivation of the formula}
In the original bases, the Green function $G_{ij}(z,\bm{B})$ is defined through the equation
\begin{equation}
  \sum_j\mathscr{L}_{ij}(z)e^{i\frac{q}{\hbar}A_{\bm{r}_i\bm{r}_j}}G_{jk}(z,\bm{B})=\delta_{ik} \label{eq:master}
\end{equation}
with 
\begin{equation}
 \mathscr{L}_{ij}(z)=z\delta_{ij}-H^{(0)}_{ij}.
\end{equation}
Our goal is to expand the Green function $G_{ij}(z,\bm{B})$ up to the second order in $B$. In the following, we abbreviate $G_{jk}(z,\bm{B})$ as $G_{jk}$ and $\mathscr{L}_{ij}(z)$ as $\mathscr{L}_{ij}$.  

Now, we introduce an ansatz,
\begin{equation}
 G_{jk} = \sum_{\alpha\beta}\tilde{w}_{j\alpha}\tilde{G}_{\alpha\beta}\tilde{w}^*_{k\beta},
\end{equation}
with
\begin{equation}
 \tilde{w}_{i\alpha}=e^{i\frac{q}{\hbar}A_{\bm{r}_i\bar{\bm{r}}_\alpha}}w_{i\alpha}.
\end{equation}
Here, $w_{i\alpha}$ is a Wannier orbital indexed by $\alpha$, and 
$\bar{\bm{r}}_\alpha$ is a Wannier center, i.e.,
\begin{equation}
 \bar{\bm{r}}_{\alpha} = \langle w_\alpha|r|w_\alpha\rangle = \sum_i \bm{r}_iw^*_{i\alpha} w_{i\alpha}. 
\end{equation}
Using the ansatz, we have
\begin{equation}
 \mathrm{Tr}G = \sum_{i}\sum_{\alpha\beta} \tilde{w}_{i\alpha}\tilde{G}_{\alpha\beta}\tilde{w}^*_{i\beta} = \mathrm{Tr}\tilde{G}\tilde{S} = \mathrm{Tr}\tilde{\mathcal{G}}
\end{equation}
with 
\begin{equation}
 \tilde{S}_{\alpha\beta}=\sum_i\tilde{w}_{i\alpha}^*\tilde{w}_{i\beta}
\end{equation}
and $\tilde{\mathcal{G}}=\tilde{G}\tilde{S}$. 
Then, Eq.~\eqref{eq:master} becomes
\begin{equation}
 \sum_j\sum_{\beta\gamma}\mathscr{L}_{ij}e^{i\frac{q}{\hbar}A_{\bm{r}_i\bm{r}_j}}
  \tilde{w}_{j\beta}\tilde{G}_{\beta\gamma}\tilde{w}^*_{k\gamma}
  =\delta_{ik},
\end{equation}
leading to
\begin{equation}
 \sum_{ijk\beta\gamma}\tilde{w}^*_{i\alpha}\mathscr{L}_{ij}e^{i\frac{q}{\hbar}A_{\bm{r}_i\bm{r}_j}}
  \tilde{w}_{j\beta}\tilde{G}_{\beta\gamma}\tilde{w}^*_{k\gamma}\tilde{w}_{k\delta}
  =\sum_{i}\tilde{w}_{i\alpha}^*\tilde{w}_{i\delta}.
\end{equation}
This can be rewritten as
\begin{equation}
 \sum_{ij\beta}w^*_{i\alpha}\mathscr{L}_{ij}e^{i\frac{q}{\hbar}\Phi_{\bar{\bm{r}}_\alpha\bm{r}_i\bm{r}_j\bar{\bm{r}}_\beta}}
  w_{j\beta}e^{i\frac{q}{\hbar}\Phi_{\bar{\bm{r}}_\alpha\bar{\bm{r}}_\beta\bar{\bm{r}}_\gamma}}\mathcal{G}_{\beta\gamma}
  =S_{\alpha\gamma}, \label{eq:transformed}
\end{equation}
where $\mathcal{G}$ and $S$ are introduced as
\begin{equation}
 \tilde{\mathcal{G}}_{\beta\gamma}=e^{i\frac{q}{\hbar}A_{\bar{\bm{r}}_\beta\bar{\bm{r}}_\gamma}}\mathcal{G}_{\beta\gamma},\quad
   \tilde{S}_{\alpha\gamma}=e^{i\frac{q}{\hbar}A_{\bar{\bm{r}}_\alpha\bar{\bm{r}}_\gamma}}S_{\alpha\gamma},
\end{equation}
and 
\begin{align}
 \Phi_{\bar{\bm{r}}_\alpha\bm{r}_i\bm{r}_j\bar{\bm{r}}_\beta} &= A_{\bar{\bm{r}}_\alpha\bm{r}_i}+A_{\bm{r}_i\bm{r}_j}+A_{\bm{r}_j\bar{\bm{r}}_\beta}+A_{\bar{\bm{r}}_\beta\bar{\bm{r}}_\alpha},\\
  \Phi_{\bar{\bm{r}}_\alpha\bar{\bm{r}}_\beta\bar{\bm{r}}_\gamma} &= A_{\bar{\bm{r}}_\alpha\bar{\bm{r}}_\beta}+A_{\bar{\bm{r}}_\beta\bar{\bm{r}}_\gamma}+A_{\bar{\bm{r}}_\gamma\bar{\bm{r}}_\alpha}.
\end{align}
Note that $\mathrm{Tr}\tilde{\mathcal{G}}=\mathrm{Tr}\mathcal{G}$ because of $A_{\bm{r}_\alpha\bm{r}_\alpha}=0$. 
Very importantly, both of $\Phi_{\bar{\bm{r}}_\alpha\bm{r}_i\bm{r}_j\bar{\bm{r}}_\beta}$ and $\Phi_{\bar{\bm{r}}_\alpha\bar{\bm{r}}_\beta\bar{\bm{r}}_\gamma}$ are obtained by line integrating $\bm{A}(\bm{r})$ on a closed path, and therefore, gauge invariant and linear in $B_\mu$ \cite{PhysRevB.84.205137}. 

Equation.~\eqref{eq:transformed} can further be rewritten as
\begin{equation}
 \sum_{\beta} (zS_{\alpha\beta}-H_{\alpha\beta})\mathcal{G}_{\beta\gamma}
  e^{i\frac{q}{\hbar}\Phi_{\bar{\bm{r}}_\alpha\bar{\bm{r}}_\beta\bar{\bm{r}}_\gamma}}=S_{\alpha\gamma}, \label{eq:gauge_master}
\end{equation}
with
\begin{equation}
 H_{\alpha\beta} = \sum_{ij}e^{i\frac{q}{\hbar}\Phi_{\bar{\bm{r}}_\alpha\bm{r}_i\bm{r}_j\bar{\bm{r}}_\beta}}w^*_{i\alpha}H_{ij}^{(0)}w_{j\beta}.
\end{equation}
Note that $H_{\alpha\beta}$ is related to $H_0$ as we show shortly. Now, our goal is to derive $\mathcal{G}$ according to Eq.~\eqref{eq:gauge_master}, and extract the terms second order in $B_\mu$. 
Up to the second order in $\Phi_{\bar{\bm{r}}_\alpha\bm{r}_i\bar{\bm{r}}_\beta}$, we have
\begin{equation}
 S\sim S_{0}+\frac{q}{\hbar}S_{1}+\frac{q^2}{\hbar^2}S_{2}
\end{equation}
with
\begin{align}
 (S_0)_{\alpha\beta} &= \delta_{\alpha\beta},\\
 (S_1)_{\alpha\beta} &= i\sum_i\Phi_{\bar{\bm{r}}_\alpha\bm{r}_{i}\bar{\bm{r}}_\beta}w^*_{i\alpha}w_{i\beta},\\
 (S_2)_{\alpha\beta} &= -\frac{1}{2}\sum_{i}\Phi_{\bar{\bm{r}}_\alpha\bm{r}_i\bar{\bm{r}}_\beta}^2w^*_{i\alpha}w_{i\beta},
\end{align}
and similarly, up to the second order in $\Phi_{\bar{\bm{r}}_\alpha\bm{r}_i\bm{r}_j\bar{\bm{r}}_\beta}$, we have
\begin{equation}
 H\sim H_{0}+\frac{q}{\hbar}H_{1}+\frac{q^2}{\hbar^2}H_{2}
\end{equation}
with
\begin{align}
 (H_0)_{\alpha\beta} &= \sum_{ij}w^*_{i\alpha}H^{(0)}_{ij}w_{j\beta}\\
 (H_1)_{\alpha\beta} &= i\sum_{ij}\Phi_{\bar{\bm{r}}_\alpha\bm{r}_i\bm{r}_j\bar{\bm{r}}_\beta}w^*_{i\alpha}H^{(0)}_{ij}w_{j\beta}\\
 (H_2)_{\alpha\beta} &= -\frac{1}{2}\sum_{ij}\Phi_{\bar{\bm{r}}_\alpha\bm{r}_i\bm{r}_j\bar{\bm{r}}_\beta}^2w^*_{i\alpha}H^{(0)}_{ij}w_{j\beta}.
\end{align}
For the later use we define $\mathcal{L}_i$ as
\begin{equation}
 \mathcal{L}_i=zS_i-H_i,
\end{equation}
and we write $\mathcal{L}_0=\mathcal{L}$ for notational simplicity. 

The heart of this formulation is to expand terms with structure of 
\begin{equation}
 \sum_\beta X_{\alpha\beta}Y_{\beta\gamma}e^{i\frac{q}{\hbar}\Phi_{\bar{\bm{r}}_\alpha\bar{\bm{r}}_\beta\bar{\bm{r}}_\gamma}}
\end{equation}
in a series of $\Phi_{\bar{\bm{r}}_\alpha\bar{\bm{r}}_\beta\bar{\bm{r}}_\gamma}$, which can also be written as
\begin{equation}
\begin{split}
  \Phi_{\bar{\bm{r}}_\alpha\bar{\bm{r}}_\beta\bar{\bm{r}}_\gamma}
  &= \frac{\bm{B}}{2}\cdot [(\bar{\bm{r}}_\alpha-\bar{\bm{r}}_\beta)\times(\bar{\bm{r}}_\gamma-\bar{\bm{r}}_\beta)]\\
  &= -\beta_{\nu\lambda}(\bar{r}_{\alpha\nu}-\bar{r}_{\beta\nu})(\bar{r}_{\beta\lambda}-\bar{r}_{\gamma\lambda})
\end{split}
\end{equation}
with $\beta_{\nu\lambda}=B_{\mu}\epsilon_{\mu\nu\lambda}/2$ and $\bar{r}_{\alpha\nu}$ being $\nu$ component of $\bar{\bm{r}}_\alpha$. The first order contribution involves a term like
\begin{equation}
 \begin{split}
 &\sum_\beta X_{\alpha\beta}Y_{\beta\gamma}\Phi_{\bar{\bm{r}}_\alpha\bar{\bm{r}}_\beta\bar{\bm{r}}_\gamma}\\
 &=-i\beta_{\nu\lambda}(\bar{r}_{\alpha\nu}-\bar{r}_{\beta\nu})(\bar{r}_{\beta\lambda}-\bar{r}_{\gamma\lambda})X_{\alpha\beta}Y_{\beta\gamma}\\
 &=\Bigl(-i\beta_{\nu\lambda}[\hat{r}_\nu,\hat{X}][\hat{r}_\lambda,\hat{Y}]\Bigr)_{\alpha\gamma}  
 \end{split}
\end{equation}
with 
\begin{equation}
 (\hat{r}_\nu)_{\alpha\beta}=\bar{r}_{\alpha\nu}\delta_{\alpha\beta}.
\end{equation}
Similarly, the second order contribution involves
\begin{multline}
 \sum_\beta X_{\alpha\beta}Y_{\beta\gamma}\Phi^2_{\bar{\bm{r}}_\alpha\bar{\bm{r}}_\beta\bar{\bm{r}}_\gamma}\\
 =
\Bigl(\beta_{\nu\lambda}\beta_{\nu'\lambda'}
  [\hat{r}_\nu,[\hat{r}_{\nu'},\hat{X}]][\hat{r}_{\lambda},[\hat{r}_{\lambda'},\hat{Y}]]\Bigr)_{\alpha\gamma}.
\end{multline}

Now, approximating $\mathcal{G}$ as 
\begin{equation}
 \mathcal{G}\sim g + \frac{q}{\hbar}g_1+\frac{q^2}{\hbar^2}g_2
\end{equation}
where $g_i$ is in the $i$th order in $B_\mu$ and $g\equiv g_0$. In the zeroth order in $B_\mu$, Eq.~\eqref{eq:gauge_master} leads to
\begin{equation}
 \mathcal{L}g=1.
\end{equation}
In the first order in $B_\mu$, Eq.~\eqref{eq:gauge_master} gives
\begin{equation}
 \mathcal{L}_1g+\mathcal{L}g_1-i\beta_{\nu\lambda}[\hat{r}_\nu,\mathcal{L}][\hat{r}_\lambda,g]=S_1,
\end{equation}
leading to
\begin{equation}
 g_1 = g\tilde{L}g+i\beta_{\nu\lambda}g\gamma_\nu g^\lambda,
\label{eq:appendix_g1}
\end{equation}
where $\gamma_\nu$, $g^\lambda$, and $\tilde{L}$ are introduced as
\begin{equation}
 \gamma_\nu=i[\hat{r}_\nu,H_0],\quad
  g^\lambda=i[\hat{r}_\lambda,g],
\end{equation}
and
\begin{equation}
 \tilde{L}=S_1\mathcal{L}-\mathcal{L}_1=H_1-S_1H_0.
\end{equation}
In the second order in $B_\mu$, Eq.~\eqref{eq:gauge_master} reduces to
\begin{multline}
 \mathcal{L}_2g+\mathcal{L}g_2-\frac{1}{2}\beta_{\nu\lambda}\beta_{\nu'\lambda'}
  [\hat{r}_\nu,[\hat{r}_{\nu'},\mathcal{L}]][\hat{r}_{\lambda},[\hat{r}_{\lambda'},g]]\\
  +\mathcal{L}_1g_1-i\beta_{\nu\lambda}[\hat{r}_\nu,\mathcal{L}_1][\hat{r}_\lambda,g]\\
  -i\beta_{\nu\lambda}[\hat{r}_\nu,\mathcal{L}][\hat{r}_\lambda,g_1]=S_2,
\end{multline}
giving us
\begin{multline}
 g_2 = g\tilde{M}g-\frac{1}{2}\beta_{\nu\lambda}\beta_{\nu'\lambda'}g\gamma_{\nu\nu'}g^{\lambda\lambda'}-g\mathcal{L}_1g_1\\
 +\beta_{\nu\lambda}g[\hat{r}_\nu,\mathcal{L}_1]g^{\lambda}
 -\beta_{\nu\lambda}g\gamma_\nu[\hat{r}_\lambda,g_1],
\label{eq:appendix_g2}
\end{multline}
where $\gamma_{\nu\nu'}$, $g^{\lambda\lambda'}$, and $\tilde{M}$ are introduced as
\begin{equation}
 \gamma_{\nu\nu'}=-[\hat{r}_\nu,[\hat{r}_{\nu'},H_0]],\quad
  g^{\lambda\lambda'}=-[\hat{r}_\lambda,[\hat{r}_{\lambda'},g]],
\end{equation}
and
\begin{equation}
 \tilde{M}=S_2\mathcal{L}-\mathcal{L}_2=H_2-S_2H_0.
\end{equation}
By subsituting $g_1$ in Eq.~(\ref{eq:appendix_g1}) into Eq.~(\ref{eq:appendix_g2}), we obtain $g_2$. For convenience, we decompose $g_2$ as
\begin{equation}
 g_2 = g_2^{(1)}+g_2^{(2)}+g_2^{(3)}+g_2^{(4)}
\end{equation}
with 
\begin{align}
 g_2^{(1)} &= g\tilde{M}g,\\
 g_2^{(2)} &= -g\mathcal{L}_1g\tilde{L}g,\\
 g_2^{(3)} &= -i\beta_{\nu\lambda}\bigl(ig[\hat{r}_\nu,\mathcal{L}_1]g^{\lambda}-ig\gamma_\nu[\hat{r}_\lambda,g\tilde{L}g]+g\mathcal{L}_1g\gamma_{\nu}g^{\lambda}\bigr),\\
 g_2^{(4)} &= \beta_{\nu\lambda}\beta_{\nu'\lambda'}
  \bigl(
  ig\gamma_\nu[\hat{r}_\lambda,g\gamma_{\nu}g^{\lambda}]
  -\frac{1}{2}g\gamma_{\nu\nu'}g^{\lambda\lambda'}
\bigr).
\end{align}
Let us rewrite these first to fourth terms $g_2^{(1)}$-$g_2^{(4)}$, respectively named Term 1 to 4,  in convenient forms in the following. 

\paragraph{Term 1} Introducing a symmetrized operator $M$ instead of $\tilde M$ as
\begin{equation}
 M=H_2-\frac{1}{2}\{S_2,H_0\}=\tilde{M}+\frac{1}{2}[S_2,\mathcal{L}],
\end{equation}
we have
\begin{equation}
 \mathrm{Tr}g_2^{(1)}=\mathrm{Tr}gMg-\frac{1}{2}\mathrm{Tr}g[S_2,\mathcal{L}]g=-\frac{\partial}{\partial z}\mathrm{Tr}Mg
\end{equation}
where the last identity follows from $\mathrm{Tr}g[S_2,\mathcal{L}]g=\mathrm{Tr}[g,S_2]=0$ and
\begin{equation}
 \frac{\partial g}{\partial z}=-g^2.
\end{equation}

\paragraph{Term 2} Introducing a symmetrized operator $L$ instead of $\tilde L$ as
\begin{equation}
 \begin{split}
  L &= H_1-\frac{1}{2}\{S_1,H_0\}\\
  &= \frac{1}{2}\{S_1,\mathcal{L}\}-\mathcal{L}_1 = \tilde{L}-\frac{1}{2}[S_1,\mathcal{L}],
 \end{split}
\end{equation}
we have
\begin{equation}
 \begin{split}
   g_2^{(2)}=&g\Bigl(L-\frac{1}{2}\{S_1,\mathcal{L}\}\Bigr)g\Bigl(L+\frac{1}{2}[S_1,\mathcal{L}]\Bigr)g\\
  =&gLgLg-\frac{1}{2}(\{S_1,g\}Lg-gL[g,S_1])\\
  &-\frac{1}{4}\{S_1,g\}\mathcal{L}[g,S_1].
 \end{split}
\end{equation}
For the first term of the right hand side, we obtain
\begin{equation}
 \begin{split}
   \mathrm{Tr}gLgLg&=
  \frac{1}{2}\mathrm{Tr}(LgLg^2+Lg^2Lg)\\
  &= -\frac{1}{2}\frac{\partial}{\partial z}\mathrm{Tr}LgLg.
 \end{split}
\end{equation}
For the second term, we evaluate
\begin{equation}
 \begin{split}
 &\mathrm{Tr}(\{S_1,g\}Lg-gL[g,S_1])\\
 &=\mathrm{Tr}(g\{S_1,L\}g+S_1gLg-gLgS_1)\\
 &=-\frac{\partial}{\partial z}\mathrm{Tr}\{S_1,L\}g,
 \end{split}
\end{equation}
and for the third term, we evaluate
\begin{equation}
 \begin{split}
  &\mathrm{Tr}(\{S_1,g\}\mathcal{L}[g,S_1])\\
  &=\mathrm{Tr}(S_1gS_1+gS_1S_1-S_1^2g-gS_1\mathcal{L}S_1g)\\
  &=-\frac{1}{2}\frac{\partial}{\partial z}\mathrm{Tr}[S_1,[H_0,S_1]]g,
 \end{split}
\end{equation}
where the last line follows from
\begin{equation}
 S_1\mathcal{L}S_1=\frac{1}{2}\{S_1^2,\mathcal{L}\}-\frac{1}{2}[S_1,[S_1,\mathcal{L}]].
\end{equation}
Combining the above equations, $\mathrm{Tr}g_2^{(2)}$ becomes
\begin{equation}
 \mathrm{Tr}g_2^{(2)}=-\frac{1}{2}\frac{\partial}{\partial z}
  \mathrm{Tr}\Bigl(LgLg-\{S_1,L\}g
  -\frac{1}{4}[S_1,[H_0,S_1]]g\Bigr).
\end{equation}

\paragraph{Term 3}
Using $\mathrm{Tr}A[B,C]=\mathrm{Tr}[A,B]C$ and $\beta_{\nu\lambda}\gamma_{\nu\lambda}=\beta_{\nu\lambda}g^{\nu\lambda}=0$, we obtain
\begin{equation}
 \begin{split}
  &\mathrm{Tr}g_2^{(3)}
  =i\beta_{\nu\lambda}\mathrm{Tr}(
  g^{\nu}\mathcal{L}_1g^{\lambda}
  -g^{\lambda}\gamma_{\nu}g\tilde{L}g
  -g\mathcal{L}_1g\gamma_{\nu}g^{\lambda}
  )\\
  &=i\beta_{\nu\lambda}\mathrm{Tr}(
  g\gamma_{\lambda}gLg\gamma_{\nu}g
  +g\gamma_{\nu}g\gamma_{\lambda}gLg
  +gLg\gamma_{\nu}g\gamma_{\lambda}g
  )\\
  &-i\frac{\beta_{\nu\lambda}}{2}\mathrm{Tr}(
  g^{\lambda}\{S_1,\mathcal{L}\}g^{\nu}
  +g^{\lambda}\gamma_{\nu}[g,S_1]
  +\{S_1,g\}\gamma_{\nu}g^{\lambda}
  ).
 \end{split}
\end{equation}
For the first term of the right hand side, we evaluate
\begin{equation}
\begin{split}
  &\mathrm{Tr}(
  g\gamma_{\lambda}gLg\gamma_{\nu}g
  +g\gamma_{\nu}g\gamma_{\lambda}gLg
  +gLg\gamma_{\nu}g\gamma_{\lambda}g
  )\\
 &=\mathrm{Tr}(
  Lg\gamma_{\nu}g^2\gamma_{\lambda}g
  +Lg^2\gamma_{\nu}g\gamma_{\lambda}g
  +Lg\gamma_{\nu}g\gamma_{\lambda}g^2
  )\\
 &=-\frac{\partial}{\partial z}
 \mathrm{Tr}Lg\gamma_{\nu}g\gamma_{\lambda}g.
\end{split}
\end{equation}
For the second term, we evaluate
\begin{equation}
 \begin{split}
  &\beta_{\nu\lambda}\mathrm{Tr}(
  g^{\lambda}\{S_1,\mathcal{L}\}g^{\nu}
  +g^{\lambda}\gamma_{\nu}[g,S_1]
  +\{S_1,g\}\gamma_{\nu}g^{\lambda}
  )\\
  &=\beta_{\nu\lambda}\mathrm{Tr}(
  g^{\lambda}S_1\gamma_{\nu}g
  +g\gamma_{\lambda}S_1g^{\nu}
  -g^{\lambda}\gamma_{\nu}S_1g
  +gS_1\gamma_{\nu}g^{\lambda}
  )\\
  &=\beta_{\nu\lambda}\mathrm{Tr}(
  g^{\lambda}[S_1,\gamma_{\nu}]g
  +g[S_1,\gamma_{\nu}]g^{\lambda}
  )\\
  &=i\beta_{\nu\lambda}\frac{\partial}{\partial z}\mathrm{Tr}
  [\hat{r}_{\lambda},[S_1,\gamma_\nu]]g,
 \end{split}
\end{equation}
where the second line is from
\begin{multline}
 \beta_{\nu\lambda}\mathrm{Tr}(g^{\lambda}\gamma_{\nu}gS_1+S_1g\gamma_{\nu}\gamma^{\lambda})\\
 = \beta_{\nu\lambda}\mathrm{Tr}(S_1g\gamma_{\lambda}g\gamma_{\nu}g+S_1g\gamma_{\nu}g\gamma_{\lambda}g)=0,
\end{multline}
and the last line is from
\begin{equation}
 g^{\lambda}Xg+gXg^{\lambda}=i[\hat{r}_{\lambda},gXg]-ig[\hat{r}_{\lambda},X]g.
\end{equation}
Combining the above equations, $\mathrm{Tr}g_2^{(3)}$ becomes
\begin{equation}
 \begin{split}
  \mathrm{Tr}g_2^{(3)}=-i\frac{\partial}{\partial z}\mathrm{Tr}
  \Bigl(
  \beta_{\nu\lambda}Lg\gamma_{\nu}g\gamma_{\lambda}g
  -\frac{\beta_{\nu\lambda}}{2}\mathrm{Tr}[\gamma_{\nu},i[\hat{r}_{\lambda},S_1]]g
\Bigr).
 \end{split}
\end{equation}

\paragraph{Term 4}
For the case $\bm{B}={}^t(0,0,B_z)$, we can simplify $\mathrm{Tr}g_2^{(4)}$ following the procedure in Ref.~\citen{PhysRevB.91.085120}, as
\begin{equation}
    \mathrm{Tr}g_2^{(4)}=\frac{B_z^2}{8}\frac{\partial}{\partial z}\mathrm{Tr}\Bigl((\gamma_xg\gamma_y+\gamma_{xy})g\gamma_xg\gamma_yg+(x\leftrightarrow y)\Bigr).
\end{equation}

Collecting the all terms, we eventually obtain for $\bm{B}={}^t(0,0,B_z)$ 
\begin{equation}
 \begin{split}
 &\frac{\partial^2}{\partial B_z^2}\mathrm{Tr}g_2\\
 &=-\frac{\partial}{\partial z}\mathrm{Tr}
  \Bigl(
  M_z-\frac{i}{2}([\gamma_{x},\eta_{y}]-[\gamma_y,\eta_x])
  \Bigr)g\\
  &+\frac{\partial}{\partial z}\mathrm{Tr}
  \Bigl(
  \{S_1,L_z\}+\frac{1}{4}[S^z_1,[H_0,S^z_1]]
  \Bigr)g\\
  &-\frac{\partial}{\partial z}\mathrm{Tr}L_zgL_zg\\
  &-i\frac{\partial}{\partial z}\mathrm{Tr}L_zg(\gamma_x g\gamma_y-\gamma_y g\gamma_x)g\\
  &+\frac{1}{4}\frac{\partial}{\partial z}\mathrm{Tr}\Bigl((\gamma_xg\gamma_y+\gamma_{xy})g\gamma_xg\gamma_yg+(x\leftrightarrow y)\Bigr)
  \label{eq:appendix_Tr_final}
 \end{split}
\end{equation}
where $S^z_1$, $S^z_2$, $L_z$, and $M_z$ are introduced through
\begin{align}
 &S_1=S_1^zB_z, &S_2=\frac{1}{2}S_2^zB_z^2,\\
 &L=L_zB_z, &M=\frac{1}{2}M_zB_z^2,
\end{align}
and $\eta_\lambda$ is defined as
\begin{equation}
 \eta_\lambda = i[\hat{r}_\lambda,S^z_1].
\end{equation}

For the calculation of $\chi$, we need ${\rm Tr}\ G$ as shown in Eq.~(\ref{eq:free_energy_to_G}). Now
\begin{equation}
    {\rm Tr}\ G={\rm Tr}\ \tilde{\mathcal G} 
    ={\rm Tr}\ {\mathcal G} 
    = {\rm Tr}\left( g+\frac{q}{\hbar} g_1 +\frac{q^2}{\hbar^2} g_2\right).
\end{equation}
Therefore, Eq.~\eqref{eq:appendix_Tr_final} leads to Eq.~\eqref{eq:total_chi}.

Finally, let us check the susceptibility formula in the original basis. 
Comparing Eqs.~\eqref{eq:master} and \eqref{eq:gauge_master}, we notice that the formula in the original basis can be obtained by
\begin{equation}
 H_0 \rightarrow H^{(0)},\quad \bar{r}_{\alpha\nu}\rightarrow r_{i\nu},\quad H_{1,2}=S_{1,2}=0.
\end{equation}
This means that only the last term of Eq.~\eqref{eq:appendix_Tr_final} remains, and thus we reproduce the result in Eq.~\eqref{eq:chi_orig}.


\begin{thebibliography}{10}

\bibitem{Peierls:1933tv}
R.~Peierls, Zeitschrift f{\"u}r Physik {\bfseries 80},  763 (1933).

\bibitem{PhysRevB.91.214405}
Y.~Gao, S.~A. Yang, and Q.~Niu, Phys. Rev. B {\bfseries 91},  214405 (2015).

\bibitem{PhysRevB.94.134423}
F.~Pi\'echon, A.~Raoux, J.-N. Fuchs, and G.~Montambaux, Phys. Rev. B {\bfseries
  94},  134423 (2016).

\bibitem{Wehrli:1968uz}
L.~Wehrli, Physik der kondensierten Materie {\bfseries 8},  87 (1968).

\bibitem{doi:10.1143/JPSJ.28.570}
H.~Fukuyama and R.~Kubo, J. Phys. Soc. Jpn. {\bfseries 28},  570 (1970).

\bibitem{PhysRevB.103.115117}
S.~Suetsugu, K.~Kitagawa, T.~Kariyado, A.~W. Rost, J.~Nuss, C.~M\"uhle,
  M.~Ogata, and H.~Takagi, Phys. Rev. B {\bfseries 103},  115117 (2021).

\bibitem{PhysRevB.104.035113}
I.~Tateishi, V.~K\"onye, H.~Matsuura, and M.~Ogata, Phys. Rev. B {\bfseries
  104},  035113 (2021).

\bibitem{tada2021quantum}
Y.~Tada, arXiv:2106.04071.

\bibitem{PhysRev.89.633}
E.~N. Adams, Phys. Rev. {\bfseries 89},  633 (1953).

\bibitem{PhysRevLett.2.150}
J.~E. Hebborn and E.~H. Sondheimer, Phys. Rev. Lett. {\bfseries 2},  150
  (1959).

\bibitem{HEBBORN1960105}
J.~Hebborn and E.~Sondheimer, J. Phys. Chem. Solids {\bfseries 13},  105
  (1960).

\bibitem{PhysRev.126.1636}
E.~I. Blount, Phys. Rev. {\bfseries 126},  1636 (1962).

\bibitem{ROTH1962433}
L.~Roth, J. Phys. Chem. Solids {\bfseries 23},  433 (1962).

\bibitem{HEBBORN1964741}
J.~Hebborn, J.~Luttinger, E.~Sondheimer, and P.~Stiles, J. Phys. Chem. Solids
  {\bfseries 25},  741 (1964).

\bibitem{PhysRev.136.A803}
G.~H. Wannier and U.~N. Upadhyaya, Phys. Rev. {\bfseries 136},  A803 (1964).

\bibitem{doi:10.1143/JPSJ.20.520}
S.~Ichimaru, J. Phys. Soc. Jpn. {\bfseries 20},  520 (1965).

\bibitem{PhysRev.97.869}
J.~M. Luttinger and W.~Kohn, Phys. Rev. {\bfseries 97},  869 (1955).

\bibitem{10.1143/PTP.45.704}
H.~Fukuyama, Prog. Theor. Phys. {\bfseries 45},  704 (1971).

\bibitem{doi:10.7566/JPSJ.84.124708}
M.~Ogata and H.~Fukuyama, J. Phys. Soc. Jpn. {\bfseries 84},  124708 (2015).

\bibitem{doi:10.7566/JPSJ.85.064709}
M.~Ogata, J. Phys. Soc. Jpn. {\bfseries 85},  064709 (2016).

\bibitem{doi:10.7566/JPSJ.85.074709}
H.~Matsuura and M.~Ogata, J. Phys. Soc. Jpn. {\bfseries 85},  074709 (2016).

\bibitem{PhysRev.84.814}
J.~M. Luttinger, Phys. Rev. {\bfseries 84},  814 (1951).

\bibitem{PhysRevB.76.085425}
M.~Koshino and T.~Ando, Phys. Rev. B {\bfseries 76},  085425 (2007).

\bibitem{PhysRevB.91.085120}
A.~Raoux, F.~Pi\'echon, J.-N. Fuchs, and G.~Montambaux, Phys. Rev. B {\bfseries
  91},  085120 (2015).

\bibitem{PhysRevLett.106.045504}
G.~G\'omez-Santos and T.~Stauber, Phys. Rev. Lett. {\bfseries 106},  045504
  (2011).

\bibitem{PhysRevLett.112.026402}
A.~Raoux, M.~Morigi, J.-N. Fuchs, F.~Pi\'echon, and G.~Montambaux, Phys. Rev.
  Lett. {\bfseries 112},  026402 (2014).

\bibitem{PhysRevResearch.3.013058}
S.~Ozaki and M.~Ogata, Phys. Rev. Research {\bfseries 3},  013058 (2021).

\bibitem{PhysRevB.56.12847}
N.~Marzari and D.~Vanderbilt, Phys. Rev. B {\bfseries 56},  12847 (1997).

\bibitem{PhysRevLett.53.2449}
G.~W. Semenoff, Phys. Rev. Lett. {\bfseries 53},  2449 (1984).

\bibitem{PhysRevLett.98.186809}
C.-Y. Hou, C.~Chamon, and C.~Mudry, Phys. Rev. Lett. {\bfseries 98},  186809
  (2007).

\bibitem{Wu:2016aa}
L.-H. Wu and X.~Hu, Sci. Rep. {\bfseries 6},  24347 (2016).

\bibitem{Kariyado:2017aa}
T.~Kariyado and X.~Hu, Sci. Rep. {\bfseries 7},  16515 (2017).

\bibitem{PhysRevB.82.085106}
M.~Kargarian and G.~A. Fiete, Phys. Rev. B {\bfseries 82},  085106 (2010).

\bibitem{PhysRevB.90.085132}
T.~Kariyado and Y.~Hatsugai, Phys. Rev. B {\bfseries 90},  085132 (2014).

\bibitem{BLOUNT1962305}
E.~Blount, Formalisms of Band Theory, Vol.~13 of {\em Solid State Physics}, pp.
  305--373. Academic Press, 1962.

\bibitem{PhysRevB.84.205137}
K.-T. Chen and P.~A. Lee, Phys. Rev. B {\bfseries 84},  205137 (2011).

\end{thebibliography}
\end{document}